# Assurance for Autonomy – JPL's past research, lessons learned, and future directions

Martin S. Feather, Alessandro Pinto
Office of Safety and Mission Success
Jet Propulsion Laboratory, California Institute of Technology
Pasadena, USA
{martin.s.feather, alessandro.pinto}@jpl.nasa.gov

*Abstract*— Robotic space missions have long depended on automation, defined in the 2015 NASA Technology Roadmaps as "the automatically-controlled operation of an apparatus, process, or system using a pre-planned set of instructions (e.g., a command sequence)," to react to events when a rapid response is required. Autonomy, defined there as "the capacity of a system to achieve goals while operating independently from external control," is required when a wide variation in circumstances precludes responses being pre-planned, instead autonomy follows an on-board deliberative process to determine the situation, decide the response, and manage its execution. Autonomy is increasingly called for to support adventurous space mission concepts, as an enabling capability or as a significant enhancer of the science value that those missions can return. But if autonomy is to be allowed to control these missions' expensive assets, all parties in the lifetime of a mission, from proposers through ground control, must have high confidence that autonomy will perform as intended to keep the asset safe to (if possible) accomplish the mission objectives. The role of mission assurance is a key contributor to providing this confidence, yet assurance practices honed over decades of spaceflight have relatively little experience with autonomy. To remedy this situation, researchers in JPL's software assurance group have been involved in the development of techniques specific to the assurance of autonomy. This paper summarizes over two decades of this research, and offers a vision of where further work is needed to address open issues.

## I. Introduction

Autonomy in the context of this paper is defined as the capacity of a system to achieve goals while operating independently from external control [1]. The NASA Autonomous Systems and Robotics roadmap [2] presents several advanced robotics and spacecraft autonomy technologies that will be need in the future to enable or augment space science and exploration missions. Applications include the ability to attend the operations inside habitats and spacecraft, explore extreme environments such as deep interiors, steep slopes, and long drives, build surface infrastructure, perform multi-spacecraft science experiments, and deliver cost effective long range and continuous worksite operations. Explicit mention of autonomous capabilities can also be found in the recent Planetary Science Decadal Survey [3] to enable missions such as Endurance-A, Enceladus Orbilander, and Uranus Orbiter and Probe. To enable infusion of autonomous systems in space missions, stakeholders need to understand the risk posture which requires several new advances in the area of autonomy assurance.

For more than two decades, JPL has made use of on-board autonomy to control spacecraft. One area where autonomy is essential is fault protection. Often the urgency of a response precludes interaction with ground control to determine the situation and respond in time. Autonomy must itself protect the spacecraft in the event of faults in its hardware or software, or off-nominal conditions in its environment. Typically, this is done with monitor/response pairs, where a monitor that detects the symptoms of a fault or off-nominal condition is coupled with an appropriate response. Such spacecraft fault management was the topic of a NASA workshop held in 2008, results of which are summarized in [4]. This led to development of a draft Fault Management Handbook, [5], and the challenges of providing software assurance of fault protection were broached in a breakout session at the 2012 NASA Independent Verification and Validation (IV&V) Annual Workshop and reported in [6].

As spacecraft become more complex, and as their missions become more ambitious, fault protection's complexity increases dramatically, exacerbating the challenges of its assurance. Much of the research described in Section III has targeted this problem. Addressing the challenges of future missions such as the ones mentioned in the first paragraph, however, requires a broader range of techniques and an ambitious research and development program which we outline in Section IV.

## II. Background

### A. Software Assurance

NASA defines software assurance as "the level of confidence that software is free from vulnerabilities, either intentionally designed into the software or accidentally inserted at any time during its life cycle, and that the software functions in an intended manner" [7]. Confidence is gained through a combination of activities, including following recommendations and guidelines (e.g., architectural principles, coding standards), appropriate training of software personnel, performing reviews and inspections of software artifacts (e.g., requirements reviews), automated code generation (e.g., from statecharts), analyses (e.g., static analysis, performance analysis), and plentiful testing of many kinds (e.g., unit testing, integration testing, scenario testing, acceptance testing).

Software assurance's role is to independently guide, assess and confirm that effective plans are in place to perform the above activities, that they are executed according to plan, and that their outcomes are of adequate quality, or if not, remedies are followed to correct them accordingly. More detail is in NASA's Software Assurance Standard [8].

Within the Agency, software assurance is performed by personnel at the individual NASA facilities, including, for NASA's highest priority missions, through NASA's Independent Verification and Validation Program (IV&V). A JPL-conducted study in 2014 stated the following as a value proposition for systems assurance:

> *Systems Assurance enables more **confident** decision-making by providing independent credentialed **information** to reduce uncertainty in systems decisions that depend on quality – and thus reduces **decision risk**. [9]*

Explaining further that:

> *…providing independent credentialed information…," implies that assurance makes unbiased assessments, based on credible evidence, process, and personnel.*

*B. A Brief History of the Emergence of Autonomy Assurance*

Autonomous systems performing complex tasks have a somewhat recent history. The dream of deploying a car that drives itself really became a possibility only after the second DARPA Grand Challenge in 2005, where a team from Stanford University was able to cover 132 miles in the desert in few minutes short of 7 hours [10]. The terms "Autonomy Assurance" or "Assured Autonomy" are relatively new. NASA's Aeronautics Research Mission Directorate (ARMD) introduced the "Assured Autonomy for Aviation Transportation" thrust in 2016 [11]; their initial 20-year roadmap includes the development of technologies to design, assure, verify, validate, and test complex autonomous systems. In the same year 2016, the National Science and Technology Council published the national AI R&D strategic plan. Several strategic areas were mentioned in this report including one on "ensuring the safety and security of AI systems". The strategy identified several research areas including explainability and transparency, verification and validation, security against attacks, and long-term AI safety and value alignment. We can say that 2016 was the year where serious investment in autonomy assurance started, with several funding agencies, including the DOT, IARPA, NASF, and DARPA spawning research programs in this area. In 2018, DARPA started the "Assured Autonomy" program [12] that proposed to tackle the problem of assuring autonomous cyber-physical systems with learning-enabled components in the loop using three technologies: designing for assurance, monitoring systems at run-time, and dynamic assurance cases. In the context of the University Leadership Initiative (ULI) program, NASA started funding programs on safe autonomy in 2020. The research community has also only recently coalesced around this topic by creating dedicated forums such as the AAAI Safe AI Workshop which started in 2019, and the international conference on Assured Autonomy which started in 2022. Thus, Autonomy Assurance as a field is still in its infancy.

Clearly, verification and validation (V&V) of individual functions, such as automatic control and fault management, have been studied in the past (see Section III for a list of previous assurance work at JPL). However, as new technologies become available, such as Machine Learning, and missions become more challenging, such as exploring the sub-surface environment of Enceladus, verification and validation will need to evolve to address these.

*C. Examples of Deployment and Their Limitations*

There are examples of autonomous systems being deployed in many application domains: self-driving cars, delivery drones, factory automation robots, inspection drones, several types of assistance in healthcare and in everyday life, and defense systems. We feel, however, that either these systems have been carefully engineered to reduce as much as possible the need for real autonomy, or they don't perform as well and predictably as they should. In the case of self-driving cars, vehicles for the general public such as the different Tesla models, still "require active driver supervision" [13], while other companies such as Waymo and Cruise rely on very accurate maps of the environment [14],[15]. These approaches have enabled deployment, but dependable full autonomy still seems to be out of reach, and serious accidents occur at a rate higher than human-driven cars (even after billions of miles driven as in the case of the Tesla autopilot). The approach seems to be similar for air vehicles. Delivery drones still require oversight during flight operations, travel in scarcely populated areas, and rely on good positioning sensors. Factory automation is an example of autonomous systems deployed in well-known and highly engineered environments, and where there is a limited number of highly repetitive tasks to be performed. In these applications, the problem of understanding the environment is highly simplified either through use of fiducial markers or electronic beacons, or because tasks are so deterministic and actuators so accurate that they can just be pre-programmed. In many other applications such as healthcare, there is always a human in the loop to make the final decisions. Defense systems are good examples of an application domain where the use of autonomy has been debated from technical and ethical points of view [16], and for the time being, humans are required to make any critical decision.

Even in the case of space missions there is limited use of advanced autonomous systems. For example, the Perseverance rover can execute complex sequences of tasks autonomously including monitoring its operations. However, these sequences are planned for and verified on Earth [17].

*D. Limitations of Current Approachs to the Verification and Validation of Autonomy Systems*

One of the main reasons for a limited use of more advanced autonomy in these application domains is lack of confidence about their behavior once deployed in the wild. The ability to evaluate risks, and in general to perform a thorough verification and validation of autonomy is essential for mission-critical applications where human life is at stake, or where, as in the case of space missions, development time and cost are simply too high to fail at completing that single attempt to achieve mission goals.

As of today, testing is still the method of choice for verification and validation. This method seems to have reached its scalability limit for autonomous systems. It has been estimated [18] that a self-driving car would need to be tested for tens of billions of miles to assess, with confidence, that the rate of fatal accidents is comparable to human-driven cars. After many years in operations, the Tesla autopilot does not seem to have driven that many miles [19]. The state of a system at a certain time comprises the environment, the system's condition, the system's estimates of those, the current execution state of the plan computed by the system, and predictions of future states. In a system that learns on-line, the state includes the learned models. Given this very large state space, the uncertainty in the system's estimates, the large set of possible decisions that the system can make autonomously, all compounded by historical dependencies, testing to cover all possible situations is impossible.

Unfortunately, there is a lack of theoretical results on the generalizability of a single scenario, which forces testing to be extensive, and hence costly and time consuming. Moreover, since test and evaluation are necessarily done in the field with autonomous systems that have not been fully tested, there is always risk of serious accidents.

The academic community has long advocated for different methods to complement testing. Among these efforts, it is worth mentioning scenario generation and falsification [20], formal methods [21], and the use of assurance cases [22]. While these methods represent a good starting point, a codified assurance-driven process for autonomous systems is still lacking. Such a process could have the same positive impact that standards such as ARP 4761, ARP 4754, DO178, and DO254 have had on the safety of aerospace systems.

III. PAST RESEARCH ON AUTONOMY ASSURANCE AT JPL

This section summarizes work done by members of JPL's Software Assurance Group to address the challenges of assuring autonomy. Note its limited scope – it does not cover the regular software assurance activities conducted on flight mission software, nor the extensive work done in the engineering and test areas of JPL to address autonomy. For broader discussions, see [23] and [24]. For specific mission applications there are papers that detail their V&V, e.g., [25] for the V&V of Mars landings, [26] for the V&V of Mars rovers, etc.

*A. V&V of a Fault Tolerant System – Model Checking*

In the late 1990s, work done jointly by members of the NASA/WVU Software Research Laboratory and of JPL's Software Assurance Group explored the use of model checking to validate high-level safety properties of a fault tolerant system. The system in question was a dually redundant spacecraft controller, in which a checkpoint and rollback scheme was used to provide fault tolerance during the execution of critical control sequences. Whether this system was "automation" or "autonomy" is debatable, but it does illustrate a challenge common to assurance of many kinds of autonomy, namely a large state space for which testing alone is infeasible as the means to achieve high confidence in its correct operation. As described in [27], the state space of the fully detailed system, $2^{87}$ states, would have been too large for model checking at that time. It took manual effort, abstracting away unnecessary detail leaving behind a partial specification, to reduce to this to a state-space size (of a few hundred thousand) manageable by the SPIN model checker. Three anomalies were identified by this approach. Using a spacecraft simulator capable of taking accurate timing into account, the implementation of the fault tolerant system was used to try to recreate each of these anomalies [28]. Two of the three anomalies were confirmed to be present in the implementation, while the third was determined to be impossible because of timing considerations.

*B. Remote Agent Experiment (RAX) on Deep Space 1 (DS1) - Testing and a Test Oracle*

The Remote Agent (RA), described as "the first Artificial Intelligence based closed loop autonomous control system to take control of a spacecraft" [29], flew as an experiment (RAX) on Deep Space 1 (DS1). RA combined general purpose reasoning engines, namely an on-board planner-scheduler, a robust multithreaded executive, and a model-based fault diagnosis and recovery system. These were informed by mission-specific domain models. For example, a model of the spacecraft included constraints on its safe operation, so that the planner-scheduler would know to avoid violations of those constraints. Declarative models of the spacecraft's components informed the fault diagnosis system so that it could determine the health status of those components from knowledge of the commands and observations of the spacecraft sensors.

Validation and verification of the Remote Agent is described in [30], where it is explained why traditional flight software testing approaches were felt insufficient for testing of RA's kind of autonomous software, primarily because of the explosion of possible execution paths precluding exhaustive testing. Scenario-based testing was the approach followed to gain confidence in the correct operation of RA, key to which was the careful selection of a manageable number of test cases with which to test each reasoning engine together with its model.

In a collaboration between the planner expert and software assurance, automation was introduced into the testing process to alleviate the effort it would take to do the manageable but nevertheless onerous checking of the challenging number of tests deemed necessary for testing of the planner-scheduler [31]. The automation took as input the set of spacecraft constraints given to the planner-scheduler, and generated a test oracle that, given the output of the planner-scheduler (a plan) would automatically check that the plan abided by all the input constraints. This approach followed the recommendations of [32], that test oracles be derived from specifications to help make testing reliable and cost-effective. In the case of the RA's planner-scheduler, the set of spacecraft constraints evolved over time in concert with increased understanding of the DS1 spacecraft design, so the automated generation of a test oracle helped the testing effort keep up with development. Another aspect of this work was checking that the planner-scheduler was generating valid results "for the right reasons". This was possible because the generated plans contained both a schedule of activities, and a trace of the constraints considered in their scheduling. The test oracle checked the validity of a generated plan, and that every constraint relevant to that plan was included in the trace.

*C. Fault Protection Engine – Test Case Generation*

Another effort that sought to improve the efficiency of testing of a reusable autonomous component was focused on the Fault Protection Engine (FPE), a reusable non-mission specific subsystem responsible for the management of all system level fault identification and recovery; it was used for fault protection on the Deep Impact spacecraft. See [33] for a description of FPE and the application of run-time monitoring to its verification. The software assurance work, done in collaboration with the Naval Research Laboratory, was to automate the generation of test cases [34]. The FPE's behavior was specified in Stateflow® diagrams, a graphical language from MathWorks®. These diagrams were systematically translated into SCR (Software Cost Reduction) specifications in the SCR toolset [35]. The goal of this work was to generate test cases (a set of test sequences) in a manner that would "cover" all possible system executions described by the specification. Furthermore, the SCR specification was executable, allowing for automatic checking of properties such as absence of unwanted non-determinism or circular definitions, for automatic generation of a simulation of the FPE (useful for validation purposes by demonstrating FPR's behavior), and potentially property verification using model checking.

*D. Model-Based Fault Diagnosis – Verification, Validation and Performance*

Model-based fault diagnosis (MBFD) is an approach to diagnosis of hardware failures based on a model of how the system being monitored will function. By comparing the system's *actual* behavior as seen from sensors and knowledge of the commands given to the system to the *modeled* behavior, faults can be detected. This is done with a reasoning engine to do the detection and diagnosis, informed by a model of the spacecraft. The reasoning engine remains the same from one application to another, while the model is specific to the spacecraft. MBFD had been demonstrated on the DS1 mission mentioned above, using Livingstone, a model-based system to both detect faults and direct the actions to recover from them [36]. The next version of Livingstone, called Livingstone 2, was then flown on the Earth Observing One (EO-1) satellite in a lengthy flight experiment. [37] explains how Livingstone 2's model of the satellite used a qualitative representation, and its reasoning engine used conflict-directed best-first search to identify the component modes (including failure modes) that would explain the observed states and transitions.

In the 2017-2019 timeframe, assurance of MBFD software was the topic of a multi-year assurance study performed in collaboration with the developers of an MBFD implementation. The assurance effort, described in [38] and [39], investigated three aspects of MBFD's V&V: correctness, completeness, and performance. Because the reasoning engine is independent of the spacecraft model, the engine's correctness and completeness, and a spacecraft's model's correctness and completeness, could be checked independently of one another. For performance however, it is necessary to check the engine in combination with the spacecraft model.

The MBFD capability used in this study was the Model-based Off-Nominal State Identification and Detection (MONSID) system [40]. MONSID requires a quantitative model of the monitored system's nominal behavior. A fault in the system would cause the actual behavior observed to be in contradiction with the nominal behavior, and in such a case, MONSID uses constraint suspension to identify the faulty component, or if identification cannot be narrowed down to a single component, a set of components, one of which is at fault.

In its first phase, reported in [38], the study began by using information about the design of the Soil Moisture Active Passive (SMAP) mission's spacecraft [41] and [42], which had launched a few years earlier. In particular, the study chose to model SMAP's fault protection as it related to the spacecraft's Guidance, Navigation, and Control (GNC). This choice was motivated by the relevance of fault protection and GNC to many missions. The study used SMAP hardware requirements and design information from which to build a MONSID model. The verification methods (procedures and tests) used for SMAP's hardware were applied to V&V of the MONSID model. Hardware acceptance tests were available and used in this verification. A conclusion from this work was demonstration of "the feasibility of adapting test procedures and results used to evaluate flight mission hardware functionality and behavior for use in determining the correctness of diagnostic models developed from flight unit specifications and designs". However, it was pointed out that details of expected results about a component's off-nominal behavior were not readily available, which was attributed to the reluctance to test flight units under off-nominal conditions, lest doing so would cause physical damage to those units.

In its second phase, reported in [39], the study had three thrusts: (1) to formalize the notions of the correctness and completeness of the models of MBFD, (2) to address the evaluation of performance of a diagnosis system, and (3) to consider the software development practices appropriate for the reasoning engine.

(1) Formalization of models was done by representing them as sets of logical sentences that specified the modeled system's behavior, observations, and structural properties, in the style of [43]. The study then constructed scenarios to illustrate the various ways in which models can be (in)correct and/or (in)complete.

(2) Evaluation of performance of the reasoning engine and model was done empirically, by measuring runtime during MONSID's execution. This was done on three families of models, each family having a different topology (of the model's components and how they were connected). Models of each family were scalable, systematically adding more components to lead to a larger, more complex model instance. This allowed measuring how the performance (runtime and memory usage) varied as models increased in size, as might be the case if modeling a more complex system. Regression analysis of the results pointed to the number of connections in a model as the most significant factor in determining the runtime performance.

(3) The reasoning engine was rigorously tested, using automated unit tests of the libraries used by MONSID, and tests to exercise the complete diagnosis engine in concert with a model. As tests were run, code coverage was measured. Additionally, static code analysis was performed on the code, checking four categories of properties: adherence to the set of coding rules for spacecraft code – the "Power of Ten" rules [44], absence of features that would impede maintainability, absence of unused or unnecessarily repeated code, and compliance with JPL-specific coding practices.

*E. Overall Autonomy Assurance – Assurance Cases, Efficient Testing, and Hazard Analysis*

A look at means to establish the overall confidence of space mission use of autonomy was the topic of another effort in the same 2017-2019 timeframe as the previous section's work. The study looked at assuring the introduction of an autonomous component into an existing traditionally controlled system. The autonomous controller in question, MEXEC (Multi-mission EXECutive), was described in [45] as "a lightweight on-board planning and execution system that monitors spacecraft state to robustly respond to current conditions".

The study [46] developed a small assurance case to determine and justify the assurance needed for the introduction of MEXEC, which was to be allowed to command the spacecraft in place of the spacecraft's traditional sequence-based controller (the "Sequencer"). Assurance cases, also called safety cases or dependability cases, are a means to help develop, organize, and present arguments. Their typical use is to present the argument for critical properties of complex systems [47]. In this instance, they were used to argue that since the system with the traditional controller had already undergone V&V, and since MEXEC had already undergone extensive testing, the remaining area of concern was whether there could be untoward interactions between MEXEC and the existing system.

To identify such interactions, the study explored using Systems Theoretic Process Analysis (STPA) [48]. The STPA process was followed to systematically consider possible interactions between MEXEC and the spacecraft's traditional software. This led to the recognition of a potential untoward interaction should the spacecraft's fault protection software be triggered (by a fault) while MEXEC is in control, since the fault protection response would initiate its own sequence of commands, while MEXEC would continue to issue commands, with the danger of contention between these two streams of commands. When traditional spacecraft commanding is in use, fault protection's response first disables the Sequencer and only then proceeds to issue its own commands to avoid contention. But the incorporation of MEXEC added another source of commands. The remedy was to make sure that fault protection's response disables the Sequencer, cancels MEXEC's currently executing tasknet, and disables MEXEC - only then would fault protection's response be allowed to start. However, following the STPA process to scrutinize the remedy led to recognition of a potential hazard in that: if MEXEC tasknet execution takes a while to be cancelled (if, say, it is in the middle of a computationally intensive step of calculating a new schedule), then an urgent fault protection response could be significantly delayed. Fortunately, measuring the cancellation time for the worst case (most complex) tasknet showed the delay to fault protection's response to be tolerable, alleviating the concern. Had this not been the case, a code change to force MEXEC to halt immediately would have been necessary.

*F. Flight demonstrations*

The opportunity arose to demonstrate autonomy technologies on the testbed of an in-flight spacecraft, and on the spacecraft itself during its extended mission. The spacecraft, ASTERIA (Arcsecond Space Telescope Enabling Research In Astrophysics) a CubeSat in low Earth orbit, was a technology demonstration mission to conduct astrophysical measurements. Having successfully completed its primary mission and two extensions to that mission, in its third extension several autonomy experiments were performed by uploading software to the spacecraft.

Infusing the MONSID technology (section D, above) into flight was begun by building MONSID models of ASTERIA's Attitude Control System and its components: reaction wheel assemblies, inertial measurement units, torque rods, a stellar reference unit, a coarse sun sensor, and a magnetometer [49]. Testing of MONSID and these models was done using both spacecraft data, and on a hardware testbed. Nominal (fault free) telemetry data from the orbiting ASTERIA was fed to MONSID and its models to check for false positives (incorrect identification of a fault when none was present), leading to adjustments to fault detection thresholds and other model parameters. A hardware testbed, the Small Spacecraft Dynamics Testbed (SSDT) [50], was used to test MONSID and its models for false negatives (failure to identify a fault when one was present) by working with the spacecraft and testbed personnel to identify fault conditions that would be useful to test for and could be achieved via fault injection into the testbed.

Demonstrations of MEXEC was done on ASTERIA as an in-flight experiment, in which MEXEC replicated the behavior of traditional sequences, in ASTERIA's case to perform scientific observations by directing the spacecraft to point the camera at targets and taking images [51], [52]. A second demonstration was planned to be performed on ASTERIA, but the mission ended earlier than expected, and so was done on ASTERIA's testbed instead. In this second experiment, MEXEC was used to autonomously manage ASTERIA's momentum, monitoring momentum buildup, and scheduling and executing the actions necessary to dump momentum prior to its buildup triggering fault protection.

Demonstrations of the combination of MONSID, MEXEC and special-purpose autonomy (AutoNav, autonomous determination of orbital state based on optical observations of other targets), were done on the testbed after mission end [53].

In these demonstrations, faults deliberately injected into the attitude control system during lengthy taking of an image were recognized by MONSID, which then relayed health status information to MEXEC, which in turn orchestrated the AutoNav activities to process only those images taken in the correct direction and held steady throughout image exposure.

*G. Data-driven Machine Learning – Framework for Trusted AI*

Although data-driven Machine Learning (ML) has made great progress in terrestrial applications, including processing of the voluminous science data returned by space missions, those missions have been reluctant to make use of ML as an autonomous control system of their assets. An intermediate step in this direction has been to develop an autonomous component using ML on Earth, and then use that component onboard – for example, [54] reports ML-developed classifiers to identify events in hyperspectral data, and their uploading to be used onboard the Earth Observing-1 (EO-1) spacecraft.

Over a decade ago [55] identified both the allure of ML autonomy in space, and the primary sources of reluctance to do so. Briefly, those sources were: spacecraft's insufficient computational capabilities, high aversion to risk given missions' expense and infeasibility of repair, and limited or delayed communication precluding real-time human oversight or feedback. While there has been progress in increased onboard computational capabilities, for example the Qualcomm Snapdragon on the Ingenuity helicopter on Mars, the need for high assurance remains as an impediment to more uptake.

To investigate this further, JPL teamed with The Aerospace Corporation to apply the latter's *Framework for Trusted AI* to two JPL applications. The framework [56], [57] had been crafted to encourage best practices for the implementation, assessment, and control of a wide range of AI-based applications. The purpose of the joint study was to demonstrate how the framework could be tailored towards mission assurance guidance for space missions. Two research applications were studied: a ML-trained system to identify terrain types in Mars rover imagery, and an onboard system to summarize and prioritize voluminous science observations for missions with severe data downlink constraints necessitating as much as 1,000:1 data reduction. Were these products of research to be deployed, they would be mission-critical, warranting the highest levels of assurance. The collaborative study of these two projects, reported in [58], confirmed the relevance and benefit of the framework's details to scientific space mission use-cases. Outcomes included guidance for identification of five distinct stakeholder groups – the Principal Investigator and Science Team, the Funding Agency, the Flight Engineers, the Operators who would have to interact with the remote mission, and the Science Community who would have to be confident of the validity of the data returned by the mission – who all must be assured of the autonomy in question for it to be adopted and used successfully by a mission. A further outcome was the development of an updated and specialized version of the framework specific to the onboard, space-based autonomy [59].

*H. Data-driven Machine Learning – Guidance and Application*

The study described in the previous section prompted realization of a lack of guidance for assurance personnel on how to work with systems involving ML components. In contrast with traditionally controlled systems, the assurance community has not accumulated the decades of experience in the form of best practices and procedures. This led to a follow-on effort to develop such guidance, drawing from the following sources:

- NASA-STD-7009A w/ Change 1, "Standard for Models and Simulations" [60]. This provides guidance applicable to the gamut of models and simulations, and explicitly calls for accompanying information to provide detail, stating *"As the M&S disciplines employed and application areas involved are broad, the common aspects of M&S across all NASA activities are addressed. The discipline-specific details of a given M&S should be obtained from relevant recommended practices"*.

- The Aerospace Corporation's Trusted AI Framework, mentioned in the previous section.

- The University of York's "Assurance of Machine Learning in Autonomous Systems" (AMLAS) [61]. This presents a systematic process to integrate safety assurance into development of ML components. The steps of the AMLAS process go from gathering the safety requirements of the system within which the ML component resides, through to yield a safety case for that ML component in its system context. Each step's description is accompanied by a safety case *pattern* (see [62]). Instantiating these patterns with the details of the ML component and details of its development and testing leads to the safety case.

- The plethora of information to be found in the published literature and in information on the Web, e.g., [63]

This effort, described in [64], drew from the above sources to compile guidance to inform assurance personnel as they check that a project's use of ML components complies with procedures and policies (such as those in the NASA standard). A major portion of this assurance guidance took the form of pitfalls – things to watch out for, and mitigations – ways to avoid or alleviate pitfalls. This effort then applied the guidance in a retrospective examination of a safety-critical ML system in the space domain, the ML-trained component of a laser safety cutoff system for the powerful ground-based laser beacon to be used in the Deep Space Optical Communication (DSOC) experiment. This ML component had already been developed and thoroughly V&V'd [65], so served as an exemplar against which to test the guidance.

IV. A VISION FOR THE FUTURE OF AUTONOMY ASSURANCE

Previous efforts at JPL described in Section III provide a good set of lessons learned and results from the evaluation of several tools and techniques for autonomy assurance. Other efforts in the area of V&V conducted in the context of flight missions such as the Entry, Descent, and Landing function, offer good examples where V&V using a mix of traditional methods (based on testing), and formal methods (based on mathematical models) can be very effective in Autonomy Assurance. We recognize, however, that the tools developed in these previous efforts apply to specific use cases or represent point solutions that address the specific integration of some functions for targeted use cases. Our vision is for a comprehensive, general, model-based assurance-driven process for autonomous systems that codifies lessons learned from previous approaches, and that is supported by a toolchain open to the integration of previous V&V approaches as well as development of new ones. To realize this vision, further research is needed in the following areas.

*Foundational Theory of Autonomous Systems*. Autonomous systems are built today by assembling several technological innovations in sensing, machine learning, and decision making, and then testing these systems in simulation and operational environments. While testing has been common practice for more traditional control systems as well, their design has been able to leverage theoretical results from stability and performance analysis. These theoretical results are used to define the input-output relations of state estimators and control algorithms which lead to well-defined and complete software requirements. Software assurance, therefore, can focus on checking for the correct specification and implementation of such input-output functions. In the case of autonomous systems, it seems that testing is used to assess the system performance, the satisfaction of top-level requirements, as well as the correctness of the software and hardware implementations of autonomous functions. Theoretical results on performance, stability, reachability and other common properties should be developed to decouple system-level functional analysis from software and hardware assurance, which could dramatically reduce the number of tests needed to achieve a target level of confidence in the safe operation of an autonomous system.

*Complementing Existing Testing-Based Approaches with Formal Methods*. As already discussed in Section II.D, a purely testing approach to V&V of autonomous systems is impractical. Formal methods, in their many forms, can address this problem. Specifically, by formal methods we refer to model-driven automated tools capable of assessing the correctness of the system with respect to given properties. For example, the evaluation of safety and risks using a mathematical representation of assurance cases [66] (or safety cases) is a type of formal analysis, as is the assessment of the probability of reaching an unsafe state using reachability analysis. Another set of interesting tools include model-based scenario and test generation, and correct-by-construction control synthesis methods.

To enable formal methods, several related problems must also be addressed. The first problem is the *definition of a modeling methodology for requirements and components, as well as for assurance cases*. These models must retain enough details to enable verification of properties of interest, such as risk of mission failure, but they must also be abstract enough to enable efficient verification given the computational complexity of typical automated reasoning algorithms. The modeling problem for autonomous systems is particularly hard because of the complexity and uncertainty associated with their environments.

*Safe Evolution of Autonomous Systems*. Autonomous systems exploring new environments may require updates to their internal models. Such updates can be done in batches and overseen by humans (off-line learning) or can be done by the system itself (on-line learning). Considering the unique features of space assets including the inability to supervise their operations in real-time and the inability to fix potential problems, the process that updates the internal models used to make autonomous decisions must be assured, meaning that it must be possible to predict the effect of these changes in terms of performance gains and additional risks.

*Integration of methods, processes, and tools into an assurance driven systems engineering environment for autonomous systems*. Finally, assurance methods should be seamlessly integrated into the systems engineering workflow. Processes should consider assurance starting from requirements engineering and continuing through the design lifecycle. Moreover, tool integration requires connecting activities and tools that are typical in the engineering practice to formal methods. Several challenges may arise in this integration mainly related to the *usability of formal methods*. This problem includes both a natural modeling language that can be understood by domain experts, and a natural interface to interpret verification results. An example of a requirement engineering tool that goes in this direction is FRET [67]. In general, the development of models, their assessment, and the verification of properties of interest should be more efficient than they are today and more amenable to practitioners who are not necessarily formal method experts.

V. CONCLUSIONS

This paper has identified how the growing dependence of robotic space missions on onboard autonomy necessitates a corresponding advancement of the means to assure such autonomy. A retrospective examination of over two decades of studies into autonomy assurance performed by JPL's Software Assurance researchers shows use of a mix of techniques. These included model checking, various forms of test automation, formal representation, performance evaluation, assurance cases, processes, and development of guidance. These tools and methods were exercised on space mission autonomy capabilities, using mission testbeds and, where possible, on in-flight missions on which experiments and demonstrations were allowed to be performed. Based on these experiences, this paper outlines a vision for a comprehensive, general, model-based assurance-driven process for autonomous systems that codifies lessons learned from previous efforts, and that is supported by a toolchain open to the integration of previous V&V approaches as well as development of new ones.

ACKNOWLEDGMENT

The research was carried out at the Jet Propulsion Laboratory, California Institute of Technology, under a contract with the National Aeronautics and Space Administration (80NM0018D0004).